\documentclass[a4paper,11pt]{article}
\pdfoutput=1 

\usepackage{jinstpub} 
                     
\title{ML-assisted versatile approach to Calorimeter R\&D}

\author[a,1]{A. Boldyrev,\note{Corresponding author.}}
\author[a]{D. Derkach,}
\author[a,b]{F. Ratnikov}
\author[a]{and A. Shevelev}

\affiliation[a]{National Research University Higher School of Economics,\\Laboratory of Methods for Big Data Analysis, 11 Pokrovsky blvd., Moscow 109028, Russia}
\affiliation[b]{The Yandex School of Data Analysis, 11/2 Timura Frunze St., Moscow 119021, Russia}

\emailAdd{alexey.boldyrev@cern.ch}

\abstract{Advanced detector R\&D for both new and ongoing experiments in HEP requires performing computationally intensive and detailed simulations as part of the detector-design optimisation process. We propose a versatile approach to this task that is based on machine learning and can substitute the most computationally intensive steps of the process while retaining the \textsc{GEANT4} accuracy to details. The approach covers entire detector representation from the event generation to the evaluation of the physics performance. The approach allows the use of arbitrary modules arrangement, different signal and background conditions, tunable reconstruction algorithms, and desired physics performance metrics. While combined with properties of detector and electronics prototypes obtained from beam tests, the approach becomes even more versatile. We focus on the Phase II Upgrade of the LHCb Calorimeter under the requirements on operation at high luminosity. We discuss the general design of the approach and particular estimations, including spatial and energy resolution for the future LHCb Calorimeter setup at different pile-up conditions.}


\collaboration[c]{\\on behalf of the LHCb Calorimeter Upgrade group}

\proceeding{The International Conference Instrumentation for Colliding Beam Physics\\
  INSTR20\\
  Budker Institute of Nuclear Physics, and Novosibirsk State University,\\
  Novosibirsk, Russia\\
  24 -- 28 February, 2020}

\notoc
\begin{document}
\maketitle
\flushbottom

\section{Introduction}
The calorimeters are an essential part of most of the existing and developing detectors in high energy physics.
The high luminosity delivered by the collider causes a high multiplicity and hit occupancy in the calorimeter. 
To operate in such conditions, a new generation of the calorimeters are being developed. They are characterised by high granularity (increased number of channels) and by the ability to measure the time of arrival of the particles to mitigate pile-up. Besides, it is often possible to improve the physics performance of the calorimeter using advanced detector response and reconstruction techniques, including ones based on machine learning.

To obtain an optimal physics performance during the R\&D of modern experiments in HEP, the detailed \textsc{Geant4} simulation\cite{geant4} of the calorimeter is necessary.
The optimisation cycle, within the calorimeter R\&D, comprises several computationally intensive elements, such as shower development and particle transport. Processes of multi-parametric optimisation appear to be also expensive. These factors make new approaches to calorimeter development necessary. In this work, we show that machine learning allows a quick turnover for the optimisation cycle, when parameters are changed, and eliminates manual work for re-tuning the simulation and reconstruction.

\begin{figure}[htbp]
\includegraphics[width=.99\textwidth]{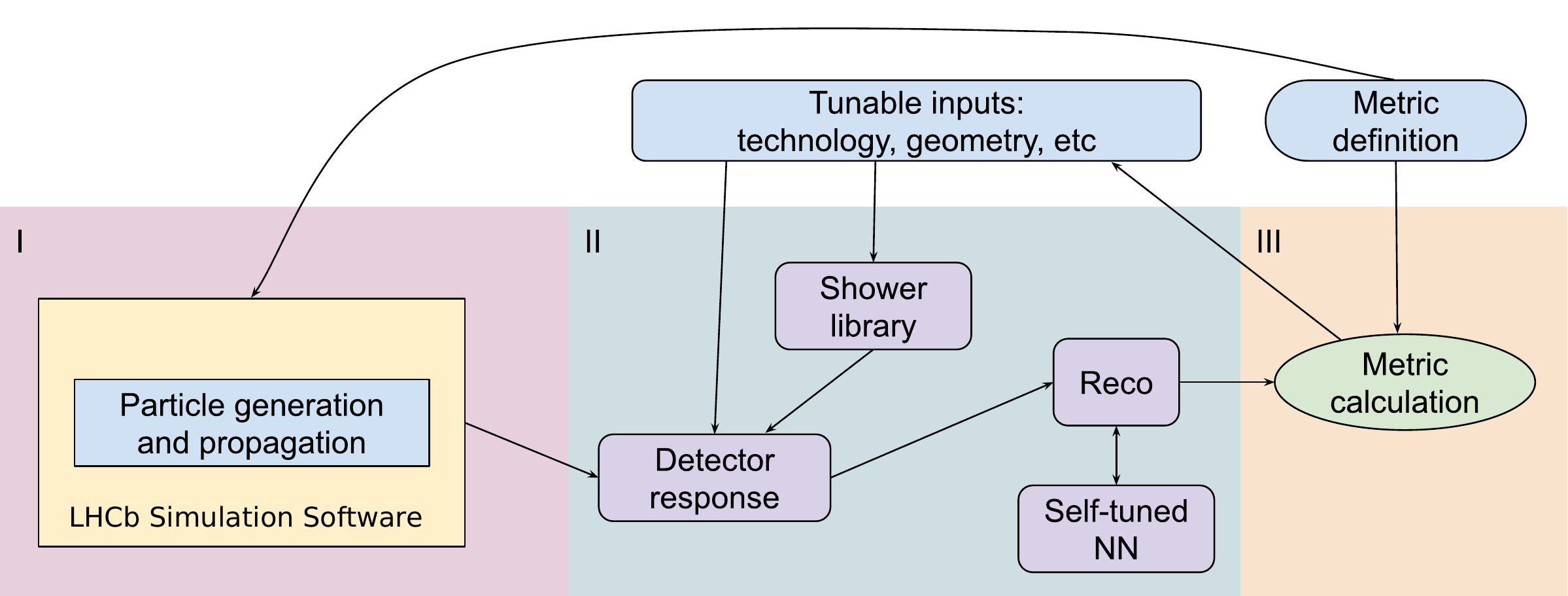}
\centering
\caption{\label{fig:pipeline} The optimisation pipeline.}
\end{figure}

We present a pipeline which optimises a calorimeter R\&D by evaluation of chosen performance metric for any interesting configuration. 
The pipeline consists of the arbitrary modules technology \& arrangement, reconstruction, metric, as depicted in figure~\ref{fig:pipeline}. 
Three main steps can represent the pipeline: particle generation and propagation (I), detector response and reconstruction (II) and metric calculation (III).

\section{Signal and background samples}
To reconstruct the hit position of any particle that reach the LHCb Electromagnetic Calorimeter (ECAL), the following simulated samples are used: $B_s^0 \rightarrow J/\psi (\rightarrow \mu^{+}\mu^{-}) \pi^0 (\rightarrow \gamma \gamma)$ (hereinafter \textit{signal sample}) and the LHCb minimum bias sample for the general detector geometry proposed to the upgrade phases (\textit{background sample}). The standalone signal sample is generated using \textsc{Pythia8} with the default LHCb tunings.

Both signal and background samples proceed through the following procedure. Each MC particle, which originated either from the primary or secondary vertices, is extrapolated to the ECAL front plane. The track extrapolation is performed using the LHCb core software framework Gaudi~\cite{Barrand:2001ny} and geometry description of the LHCb detector for the upgrade. Only track which hit positions match the ECAL outer dimensions and the inner acceptance around the beam pipe, are kept. Thus, the momentum ($p_x$,$p_y$,$p_z$), the type and the position of each particle are collected at the entrance of the calorimeter.

Our estimate is that 92\% of the contributions to the total background at the the front of the ECAL are from $\gamma, \pi^{\pm}, e^{\pm}, n, p$.
\section{Clusters generation}
 
Following the pipeline, we require to have the electromagnetic showers simulation for each type of ECAL module for signal, and for each particle contributed to the background. The output of the simulation should be uniform and consistently applied within the module type. This allows us to arrange the modules in an arbitrary way.

The electromagnetic clusters in calorimeter are produced using the \textsc{Geant4} standalone simulation. This simplified simulation setup is realised in the \textsc{Geant4} and uses the alternating 4 mm thick scintillator tiles and 2 mm thick lead plates (the \textit{Shashlik} technology) which are arranged perpendicular to the beam pipe and it consists of a matrix of 30x30 cells of size 20.2x20.2~mm$^2$ in $\eta$ -- $\phi$ plane. This allows us to emulate each type of current ECAL modules: inner, middle or outer with cells size of 40.4x40.4~mm$^2$, 60.6x60.6~mm$^2$ or 121.2x121.2~mm$^2$, respectively.
 
In this \textsc{Geant4} standalone simulation, the photons with the momenta taken from the signal sample are required to originate from the vertex located 19.8 cm in front of the calorimeter. The vertex position in the plane which is parallel to the calorimeter front plane is shifted such that the electromagnetic shower maximum lies within the area of 1.01x1.01 cm in the calorimeter front plane. This ensures that each cluster from the signal sample remains in the calorimeter sensitive volume. The particles from the background sample may come to the calorimeter at any steep angles. So, the origin vertex for them is located much closer to the calorimeter: at 0.1 cm in front of it. The output of the simulation is the array of 30x30 of energy deposits.

Thus, for 600 thousand photons from the signal sample and for 60 thousand from each of the particle contributed to the background, the electromagnetic clusters have been produced. For each photon from the signal sample, we find the closest track in \textsc{Geant4} simulated data.\footnote{The track affinity is based on distance in \textit{px, py, pz} space.
For quick nearest-neighbour lookup the 3 dimensional kd-tree is created using the package cKDTree from SciPy\cite{scipy}.}
This shapes the shower library, which is used as an input for a realistic reconstruction in the pipeline.

\section{Pile-up modelling and clusters positioning}
The design of the pipeline allows controlling the density of the background clusters in the vicinity of the signal cluster w.r.t. the module type and position of the signal cluster in the calorimeter.

The background sample contains 50 million minimum biased events with the average number of overlapped $pp$-collisions of 7.2. MC particle information is extracted from the sample for every primary vertex to have a certain number of primary vertices per event ($nPV$). The linear extrapolation of $nPV$ is used to obtain the events with arbitrary $nPV$ greater than 10. In this case, the different events are stacked the corresponding number of times. Thus, the $nPV$ range from 1 to 50 is covered. For each signal event, we consider having exactly one event from the minimum bias sample with a certain $nPV$.

\begin{figure}[htbp]
\centering 
\includegraphics[width=.99\textwidth]{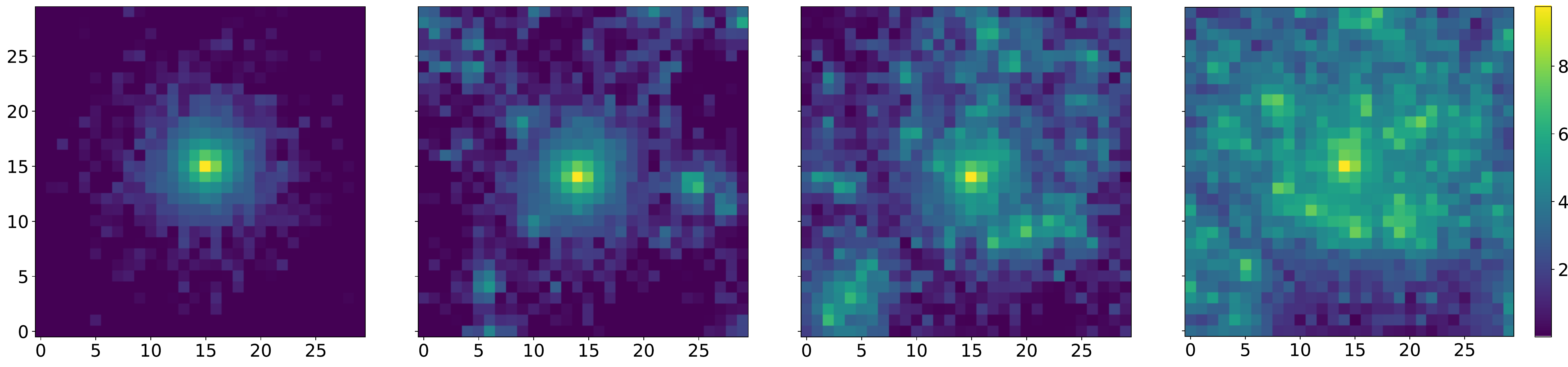}
\caption{\label{fig:clusters}The calorimetric clusters in the matrix of 30x30 cells (606x606~mm$^2$). The same signal cluster is enclosed by the background clusters for different pile-up conditions. From left to right: without pile-up, $nPV$ = 5, $nPV$ = 20, $nPV$ = 50. The colour represents $ln(E/MeV)$ for every cell.}
\end{figure}

In realistic consideration, each signal cluster is enclosed by the clusters originated from the background particles. The density of the background particles depends on the position at the calorimeter plane and on the pile-up. The different pile-up conditions for the same signal are shown in figure~\ref{fig:clusters}.

Since the cells of size 20.2x20.2~mm$^2$ is able represent each type of present ECAL modules, the initial matrix of 30x30 cells could be converted to the matrices of 15x15 for inner region modules, 10x10 for middle region modules and 5x5 for outer region modules, respectively.
The calorimeter cell, in which the signal produces the hit is required to be surrounded by two layers of cells of the same type.
Thus, a matrix of 5x5 cells of the same type is obtained.\footnote{It is assumed that the use of adjacent modules of the same type is also suitable for the borders between calorimeter regions. In this case, the adjacent module of another type on the other side of the border will be surrounded by modules of its own type. The contribution of responses of modules of different type will be compensated.} We suppose that most of the clusters of the signal sample do not exceed the size of such a matrix.
\section{Spatial reconstruction}
The reconstruction step of the pipeline uses the detector response to the signal photon under the certain pile-up conditions and the modules arrangement. It is possible to achieve adequate spatial reconstruction
by the fine-tuning of a parametric approach to the analysis of the electromagnetic clusters.
Assuming that the reconstruction does repeat for a large number of possible options listed above, this fine-tuning should be done for each combination of the options.
The pipeline allows to eliminate this manual work by using an appropriate ML-based reconstruction.
This holds true for both spatial reconstruction (this section) and energy reconstruction (section \ref{sec:energy_reco}).
Since the pipeline is able to tune the inputs of the detector response and reconstruction, the global optimisation procedure can be applied to the pipeline. This further minimises the total number of optimisation iterations.
ML proposes effective approaches to this kind of optimisation problems, like Bayesian optimisation and others\cite{bayesopt}.

\begin{figure}[htbp]
\includegraphics[width=.99\textwidth]{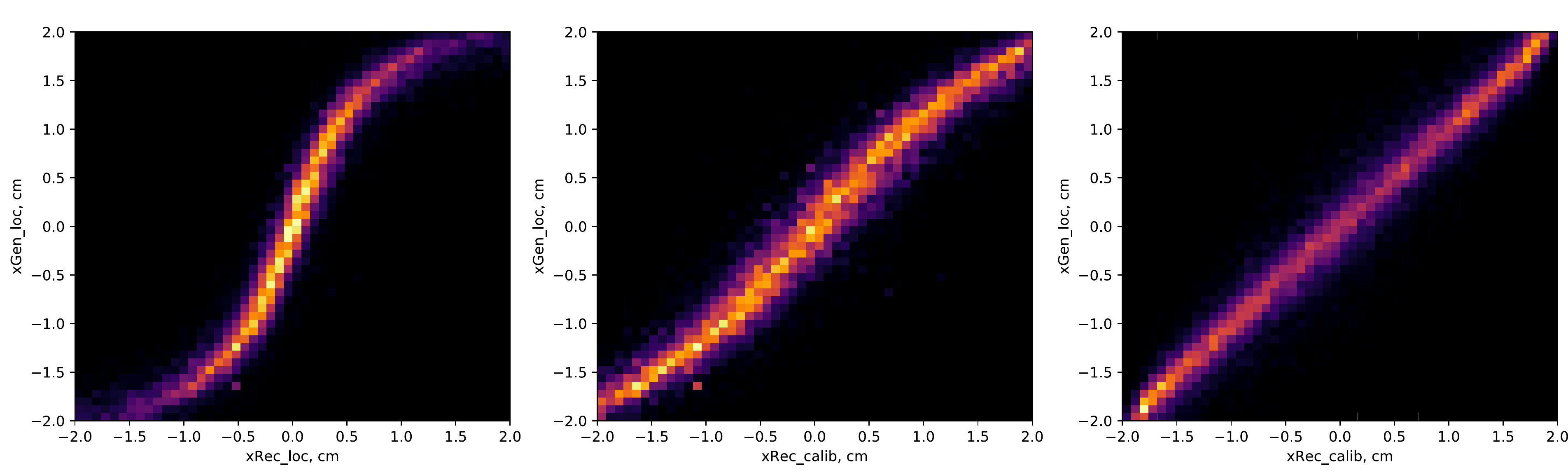}
\includegraphics[width=.99\textwidth]{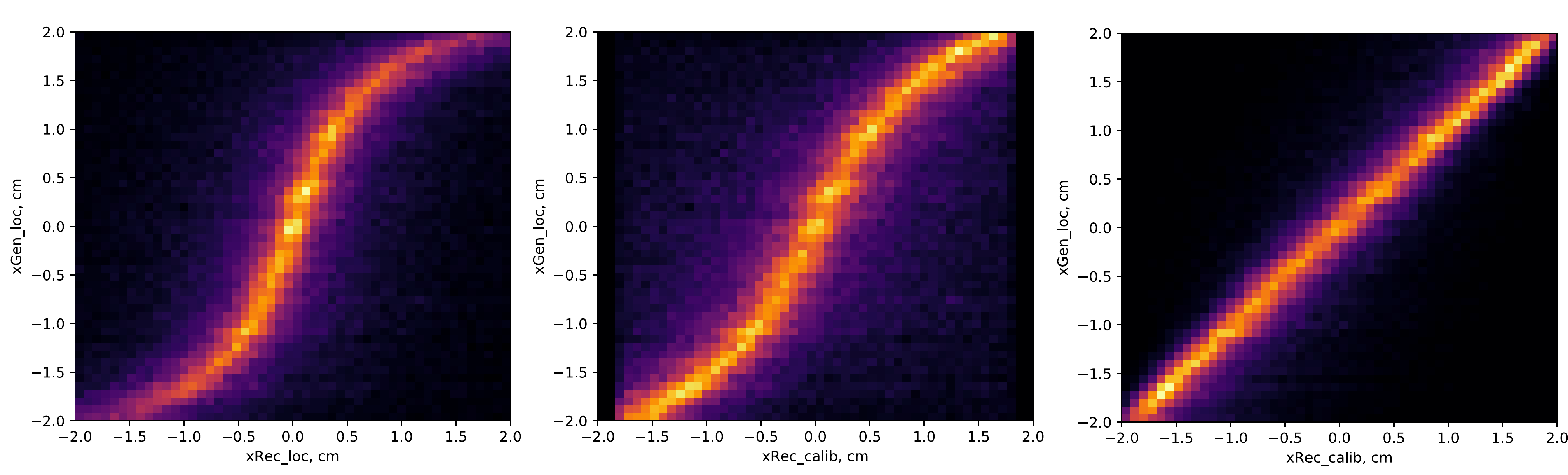}
\centering
\caption{\label{fig:S-curve_calibration} S-curve calibration results (left: uncalibrated) for inner modules and local coordinate \textit{x} using parametric approach (centre) and XGBoost regressor (right). Top row corresponds to signal in the absence of background, bottom row to $nPV$ = 10. The colour from violet (dark) to yellow (bright) represents the normalised counts of the events from 0.0 to 1.0, respectively.}
\end{figure}

The reconstructed position of the signal photon which released energy in the calorimeter is based on the barycentre of the electromagnetic cluster.
The dependence of each of the local coordinates of the signal cluster barycentre on the corresponding true coordinate of the hit position we call the S-curve due to its distinctive shape, as depicted in figure~\ref{fig:S-curve_calibration} (left).
The procedure of approaching the shape of the S-curve to a straight line we call the S-curve calibration. The calibration allows us to estimate spatial resolution uniformly both in the centre of the cell and at the borders.
Several approaches are tested to calibrate the S-curve (hit position reconstruction): the parametric approach and the machine learning approach using XGBoost regressor\cite{xgboost}. The results of the calibration for these approaches without background and for $nPV$ = 10 are shown in figure~\ref{fig:S-curve_calibration}.
To quantify a metric of spatial resolution, we use RMSE of the difference between a true and reconstructed local coordinate (independent of \textit{x} and \textit{y})\footnote{The orthogonal coordinates within the cell.} of the hit.
The observed difference in the metric, based on local coordinates \textit{x} and \textit{y} is found to be negligible. Therefore, all the results exploited in this metric are presented for local coordinate \textit{x}.

For the parametric approach, the reconstructed local coordinate \textit{x} is taken to be ${a \cdot arcsinh(b \cdot x)}$. The parameters for the calibration are found using random search\cite{random} with 1000$^2$ points in the range (0.01, 100) for each parameter. The best parameters obtained using a parametric approach for the inner section and $nPV$ = 1 (10) are: \textit{a} = 1.15 (1.13), \textit{b} = 2.07 (1.96).

The selected machine learning approach is based on the extreme gradient boosting ({\mbox{XGBoost}}) with the basic features from the observables: the barycentre position of the cluster, the particle incident angles and energy of each cell in 5x5 cells around maximum energy deposit (the seed cell).

\begin{figure}[htbp]
\includegraphics[width=.8\textwidth,trim=0 0 0 20,clip]{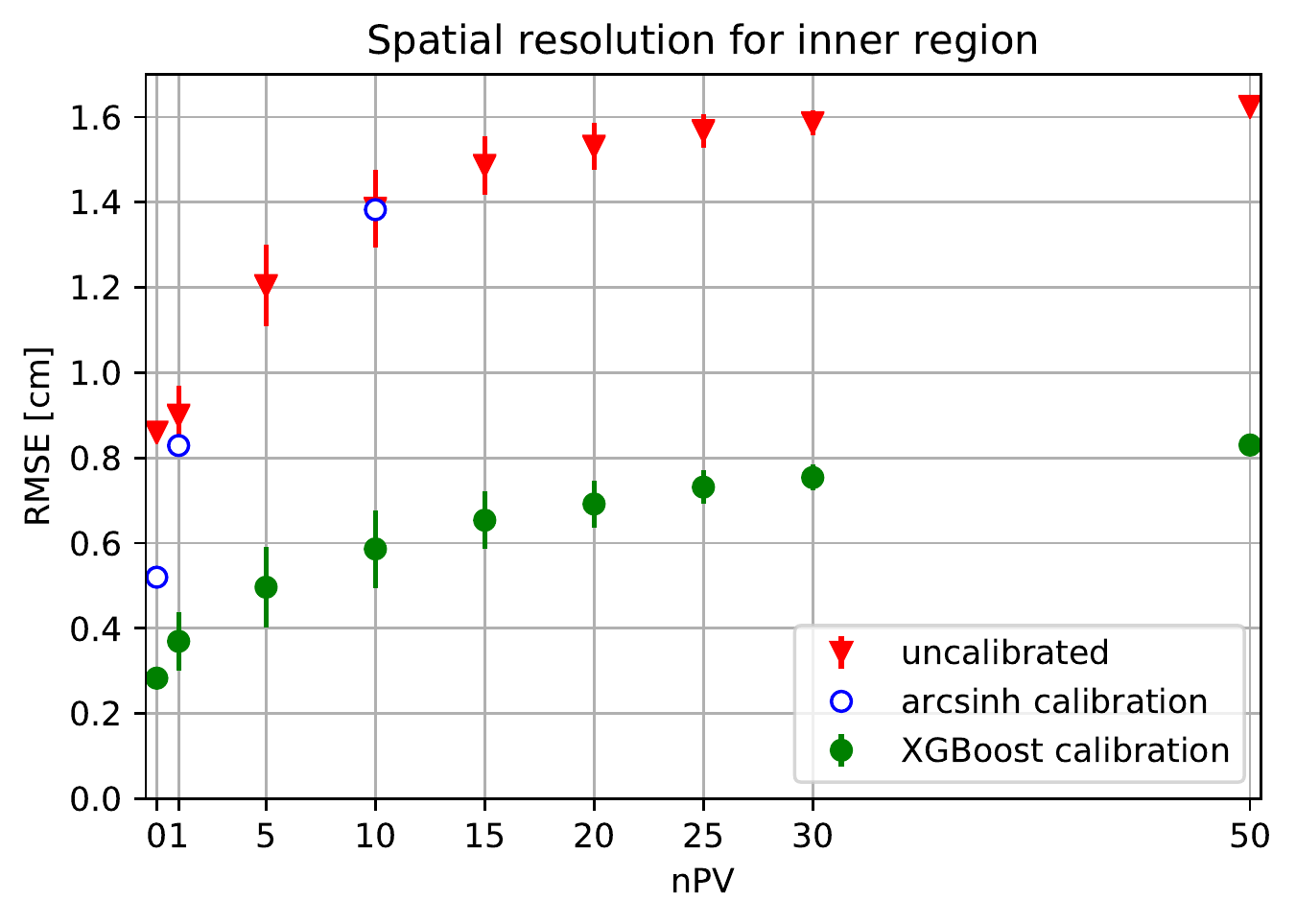}
\centering
\caption{\label{fig:inner_spatial_results} Spatial resolution as a function of $nPV$ for the ECAL inner modules (cell size 40.4x40.4 mm$^2$). $nPV$ = 0 denotes the signal without background.}
\end{figure}

The dependence of spatial resolution on $nPV$ for inner modules is shown in figure~\ref{fig:inner_spatial_results}. One can see that spatial resolution is saturated for nPV > 15. The parametric approach is acceptable for spatial reconstruction without background (denoted as $nPV$ = 0). While XGBoost shows better performance for entire range of pile-up. 
\section{\label{sec:energy_reco}Energy reconstruction}
The energy resolution of the ECAL regions had been determined using following parameterisation:\\
$$\frac{\sigma_{reco}}{E_{reco}} = \frac{a}{\sqrt{E_{gen}}} \oplus b  \oplus \frac{c}{E_{gen}},$$
where the parameters a, b and c stand for the stochastic, constant and the noise terms, respectively, and E is given in $GeV$.

\begin{figure}[htbp]
\includegraphics[width=.79\textwidth]{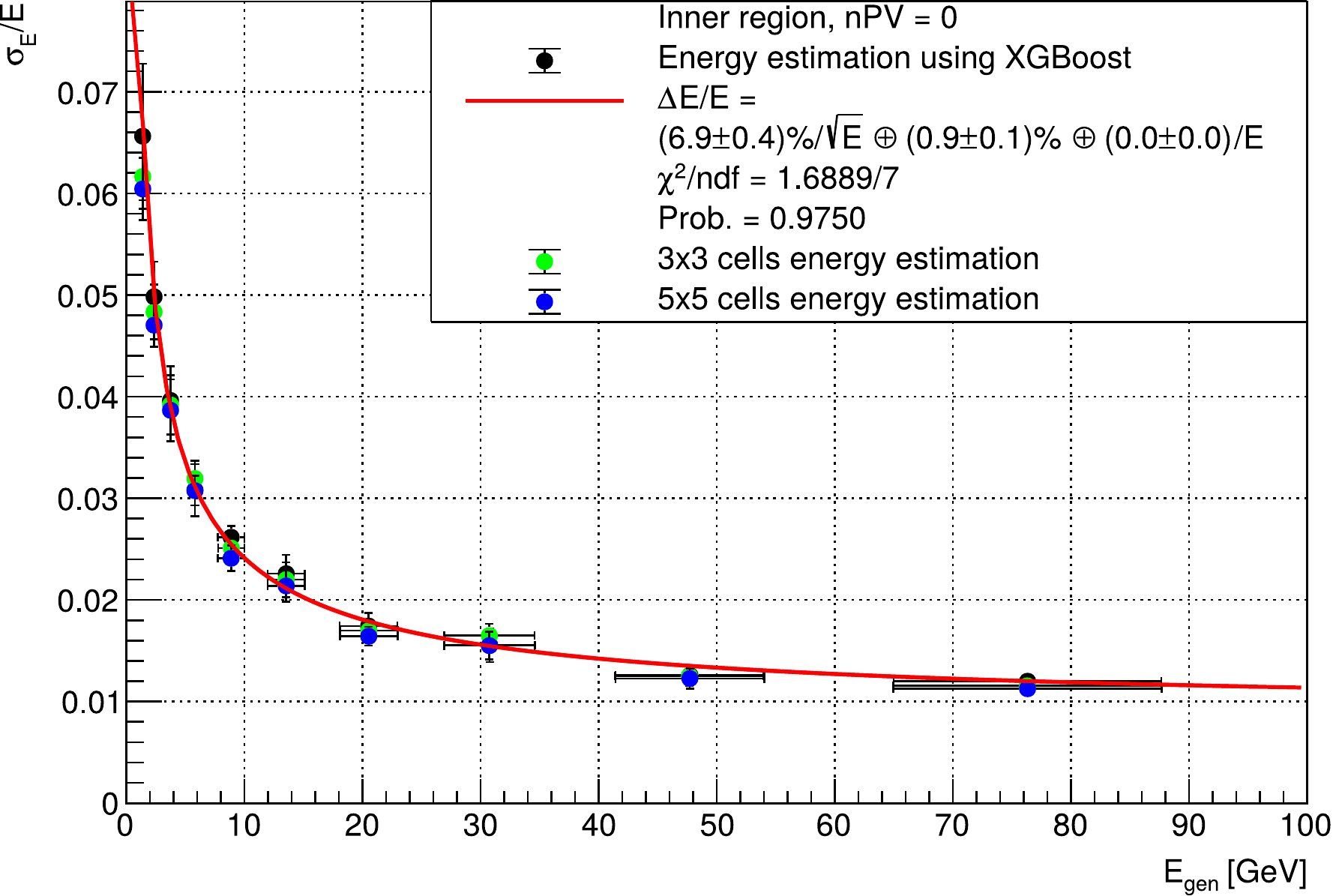}
\centering
\caption{\label{fig:energy_res_0} Energy resolution as a function of photon energy for the ECAL inner modules (signal without background). The fit (red line) corresponds to the XGBoost data (black circles). Energy is measured in GeV.}
\end{figure}

The energy is reconstructed using a machine learning approach based on XGBoost regressor such as the one we used for the spatial reconstruction. In addition, we estimate the energy using solely the sum of energy deposits in the 3x3 or 5x5 cells around the seed cell.
The XGBoost regressor uses the basic features from the observables: the barycentre position of the cluster, the particle incident angles and energy of each cell in 5x5 cells around the seed cell. The regressor is trained to minimise the difference of reconstructed energy of the signal cluster and the generated cluster energy (RMSE $\sigma_E/E$). Figure~\ref{fig:energy_res_0} shows the energy resolution for the ECAL inner modules in the absence of background.

\begin{figure}[htbp]
\includegraphics[width=.79\textwidth]{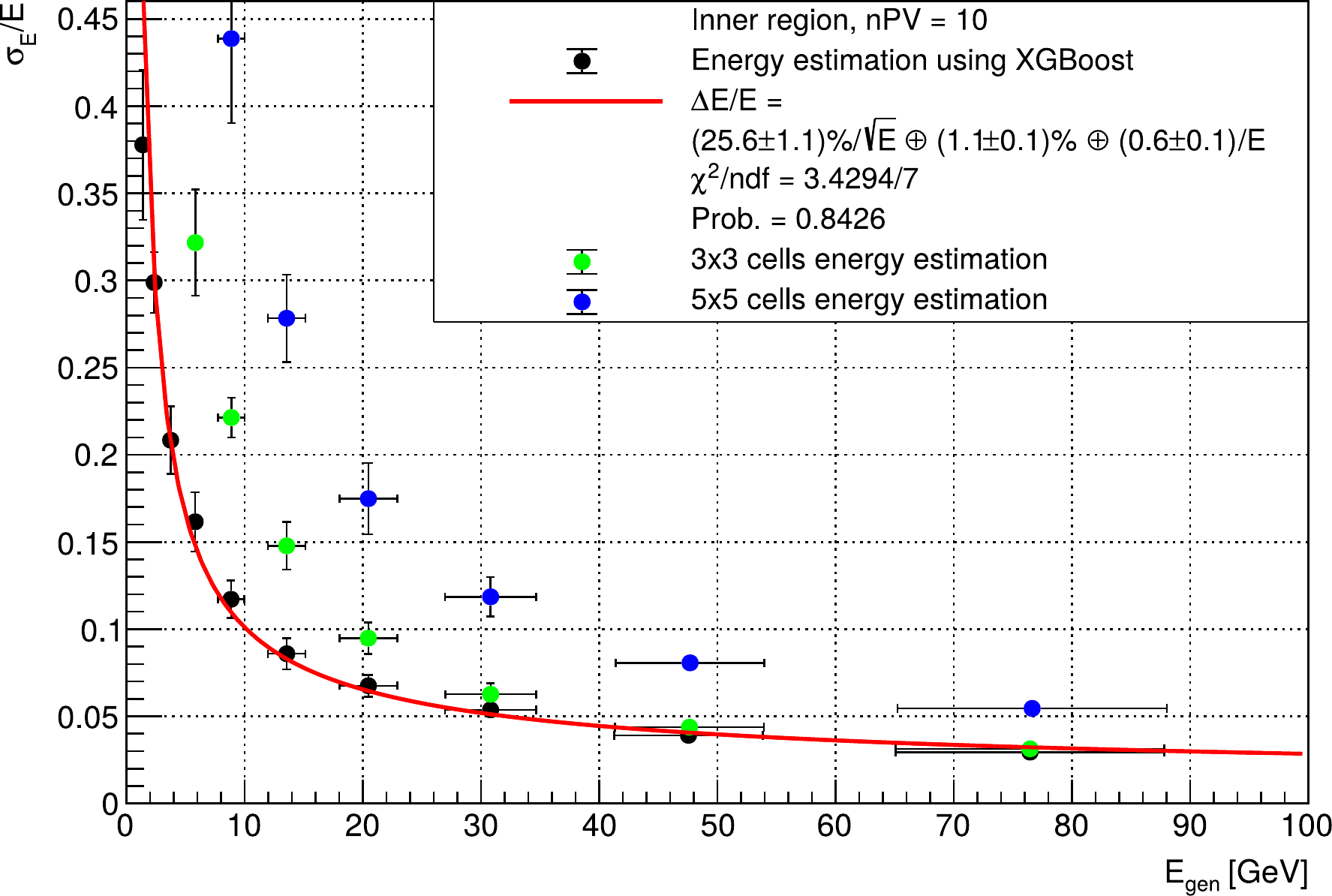}
\centering
\caption{\label{fig:energy_res_10} Energy resolution as a function of photon energy for the ECAL inner modules ($nPV$ = 10). The fit (red line) corresponds to the XGBoost data (black circles). Energy is measured in GeV.}
\end{figure}

Figure~\ref{fig:energy_res_10} shows the energy resolution for the ECAL inner modules for $nPV$ = 10. One can see that energy resolution for energy estimated using total energy in 3x3 and 5x5 cells tend to split at increased pile-up. The XGBoost regressor demonstrates better energy resolution compared to energy estimation using total energy in 3x3 cells at energies < 30~GeV. 
However, the energy resolution at $nPV$ = 10 is still insufficient for the physics measurements.

\section{Conclusions}
A pipeline for the calorimeter, using an evaluation of the chosen performance metric for any interesting module technology and configuration optimisation, is developed. Machine Learning approaches inside the pipeline substitute fine-tuning of the parameters at the simulation and reconstruction steps of the calorimeter R\&D. The pipeline can avoid most CPU-intensive parts of calorimeter full simulation while using the \textsc{Geant4} clustering. The pipeline provides physics performance for arbitrary pile-up conditions. Spatial and energy resolutions for high pile-up are presented using current LHCb electromagnetic calorimeter configuration as an example.

\acknowledgments
The research leading to these results has received funding from Russian Science Foundation under grant agreement n$^{\circ}$~19-71-30020.



\end{document}